\documentclass[article,reprint,aps,superscriptaddress,showpacs,preprintnumbers,amsmath,floatfix]{revtex4}
 
\usepackage{graphicx}

\newcommand{\be}{\begin{equation}}

\newcommand{\ee}{\end{equation}}

\newcommand{\<}{\langle}

\renewcommand{\>}{\rangle}

\begin{document}
\preprint{APS/123-QED}

\title{Observation of Scissors Modes in solid state systems with a SQUID}


\author{Keisuke Hatada}
\affiliation{
 D\'{e}partement Mat\'{e}riaux Nanosciences, Institut de Physique de Rennes UMR UR1-CNRS 6251, Universit\'{e} de Rennes 1, F-35042 Rennes Cedex, France
}
\affiliation{
 Physics Division, School of Science and Technology, Universit\`a di Camerino, Via Madonna delle Carceri 9, I-62032 Camerino (MC), Italy
}
\affiliation{INFN Laboratori Nazionali di Frascati, Via E. Fermi 40, c.p. 13, I-00044 Frascati, Italy}

\author{Kuniko Hayakawa}
\affiliation{INFN Laboratori Nazionali di Frascati, Via E. Fermi 40, c.p. 13, I-00044 Frascati, Italy}

\author{Fabrizio Palumbo}
\affiliation{INFN Laboratori Nazionali di Frascati, Via E. Fermi 40, c.p. 13, I-00044 Frascati, Italy}

\author{Augusto Marcelli}
\affiliation{INFN Laboratori Nazionali di Frascati, Via E. Fermi 40, c.p. 13, I-00044 Frascati, Italy}
\affiliation{RICMASS, Rome International Center for Materials Science Superstripes, 
00185 Rome, Italy}



\date{\today}

\begin{abstract}
ABSTRACT The occurrence of scissors modes  in crystals that have  deformed ions in their cells has been predicted some time ago. The theoretical value of their energy is rather uncertain, however, ranging between 10   and a few tenths of eV, with the   corresponding widths of $10^{-7}, 10^{-6}$ eV. Their observation  by  resonance fluorescence experiments therefore requires a photon spectrometer covering a wide energy range with a very high resolving power. We propose and discuss a new  experiment in which such difficulties are  overcome by measuring with  a SQUID the variation of the magnetic field associated with the excitation of scissors modes.
\end{abstract}

\pacs{{75.10.-b}{General theory and models of magnetic ordering},
{71.10.-w}{Theories and models of many-electron systems}}

\maketitle

\section{Introduction}

In a series of papers it has been suggested  that deformed atoms in crystal cells might have collective excitations  called scissors modes~[\onlinecite{Hata,Hata1,Hata2}]. 
These are states in which two particle systems move with respect to each other conserving their shape.
They were first predicted to occur in deformed atomic nuclei~[\onlinecite{LoIu}] by a semiclassical Two-Rotor-Model (TRM) in which  protons and neutrons were assumed to form two interacting rotors to be identified with  the blades of  scissors. Their relative motion (Fig.1) generates a magnetic dipole moment whose coupling with the electromagnetic field provides their signature.

\begin{figure}
  \begin{center}
      \includegraphics[width=3cm]{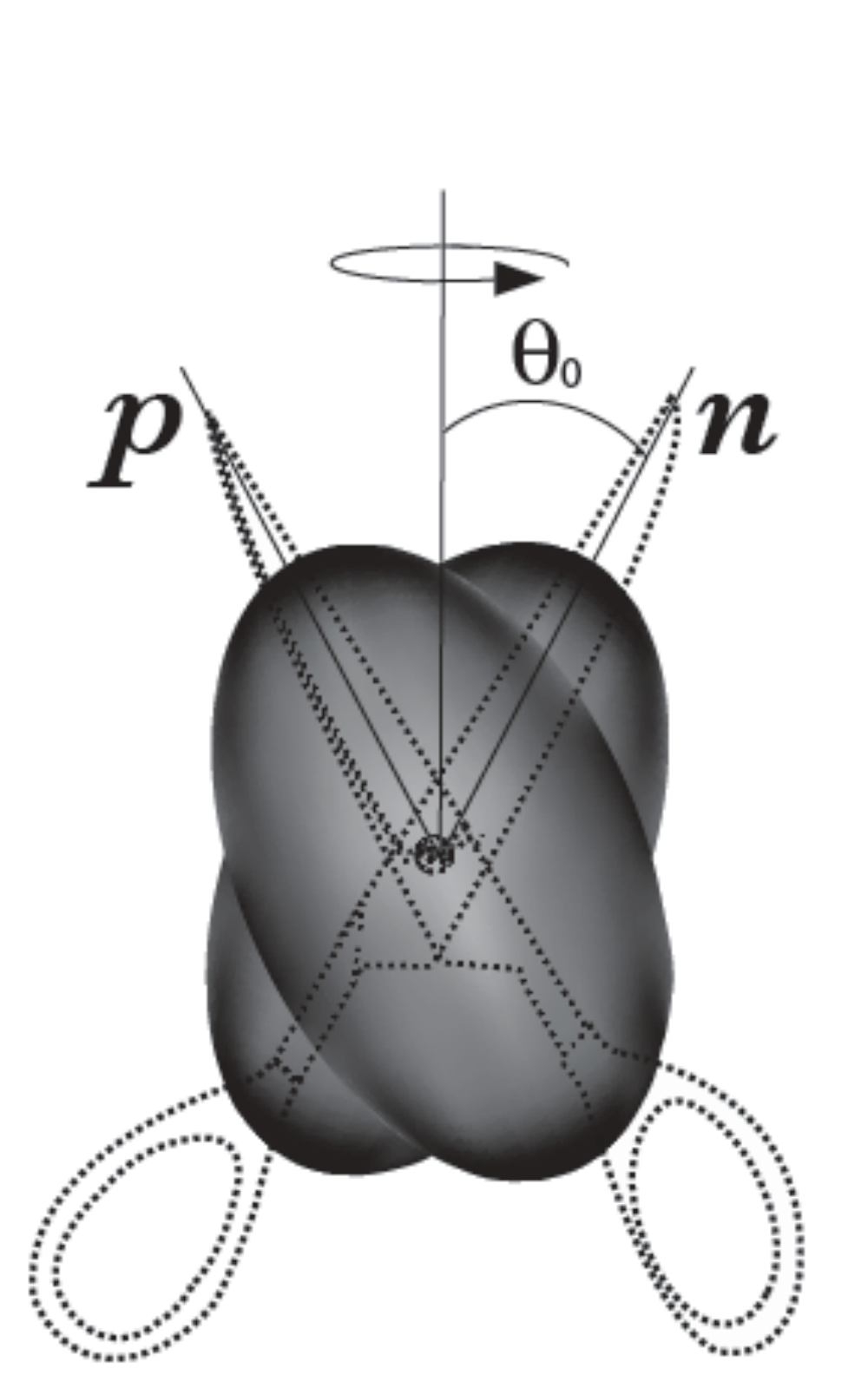}
  \end{center}
 \caption{Scissors Mode in atomic nuclei: the proton and neutron symmetry axes precess around their bisector
  }
     \label{f1}
\end{figure}

After their discovery~[\onlinecite{Bohle}] in a rare earth nucleus, $^{156}Gd $, and their systematic experimental and theoretical investigation~[\onlinecite{Ende}] in all deformed atomic nuclei, scissors modes were predicted to occur in several other systems including metal clusters~[\onlinecite{Lipp}], quantum dots~[\onlinecite{Serr}], Bose-Einstein~[\onlinecite{Guer}] and Fermi~[\onlinecite{Ming}] condensates and crystals~[\onlinecite{Hata,Hata1,Hata2}].  In all these systems one of the blades of the scissors must be identified with a moving cloud of particles (electrons in metal clusters and quantum dots, atoms in Bose-Einstein and Fermi condensates, individual atoms in   crystal cells) and the other one with a structure at rest (the trap in Bose-Einstein and Fermi condensates, the lattice in metal clusters, quantum dots and crystals). The dynamics of such  systems can then be described by a One-Rotor-Model (ORM).

More recently the TRM has been used ~[\onlinecite{Hata4}] also for  single domain free nanoparticles~[\onlinecite{Chud}]. These objects consist of a magnetic structure, called macrospin, that rotates with respect to a non magnetic lattice. They have been represented as a couple of rigid rotors, one associated with the nonmagnetic lattice, and the other one, with a spin attached, with the macrospin. If the nanoparticles are not free, but stuck in rigid matrices, the dynamics is determined by a ORM with spin, representing the macrospin.

Scissors modes have been observed not only in atomic nuclei, but also in Bose-Einstein condensates~[\onlinecite{Mara}]. Moreover 
 nanoparticles stuck in a rigid matrix were studied in detail by a ORM, and the magnetic susceptibility was found compatible with a vast body of experimental data and in some cases the agreement was surprisingly good~[\onlinecite{Hata4}]. So we can fairly state that the Two-Rotors  or the One-Rotor models are relevant to disparate physical systems involving energies and sizes at very different scales.

Scissors modes in crystals have also been studied  in the framework of a ORM in which an ion is regarded as a rigid body, which can rotate around the axes of its cell under the electrostatic force generated by the ligands. 
  We considered crystals with uniaxial~[\onlinecite{Hata}] and cubic symmetry~[\onlinecite{Hata1}]. In the first case the precessing ion was treated as one ellipsoidal rotor, in the second case as the body obtained by superimposing three ellipsoids at right angles. In the presence of uniaxial symmetry  {\it the photoabsorption cross section is characterized by a linear dichroism}~[\onlinecite{Hata}] (Fig.2) due to the fact the ion can rotate around the symmetry axis of the cell, but not  around the other ones.

   \begin{figure}
  \begin{center}
   \begin{tabular}{cc}
      \includegraphics[width=4cm]{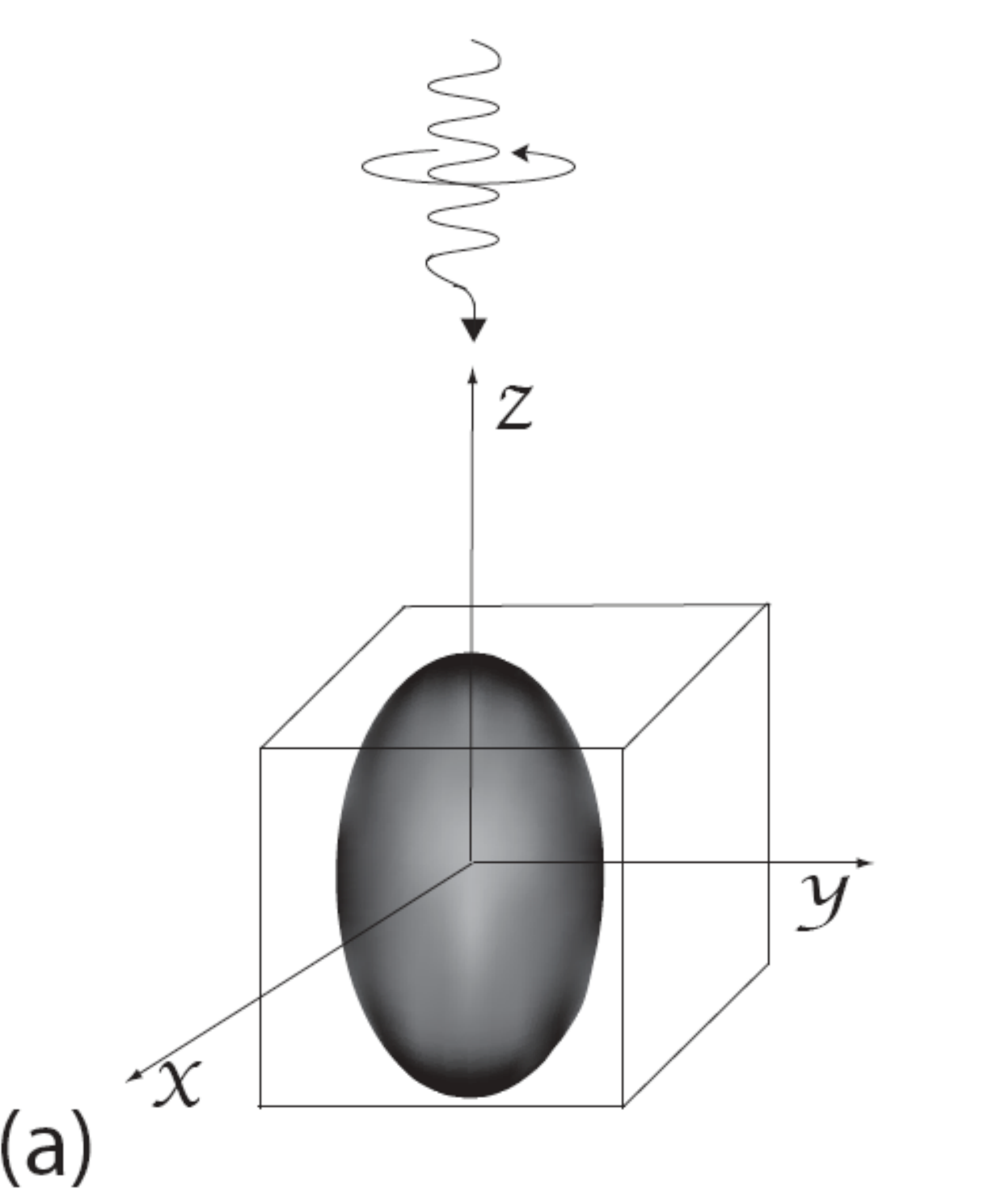}
      \includegraphics[width=4cm]{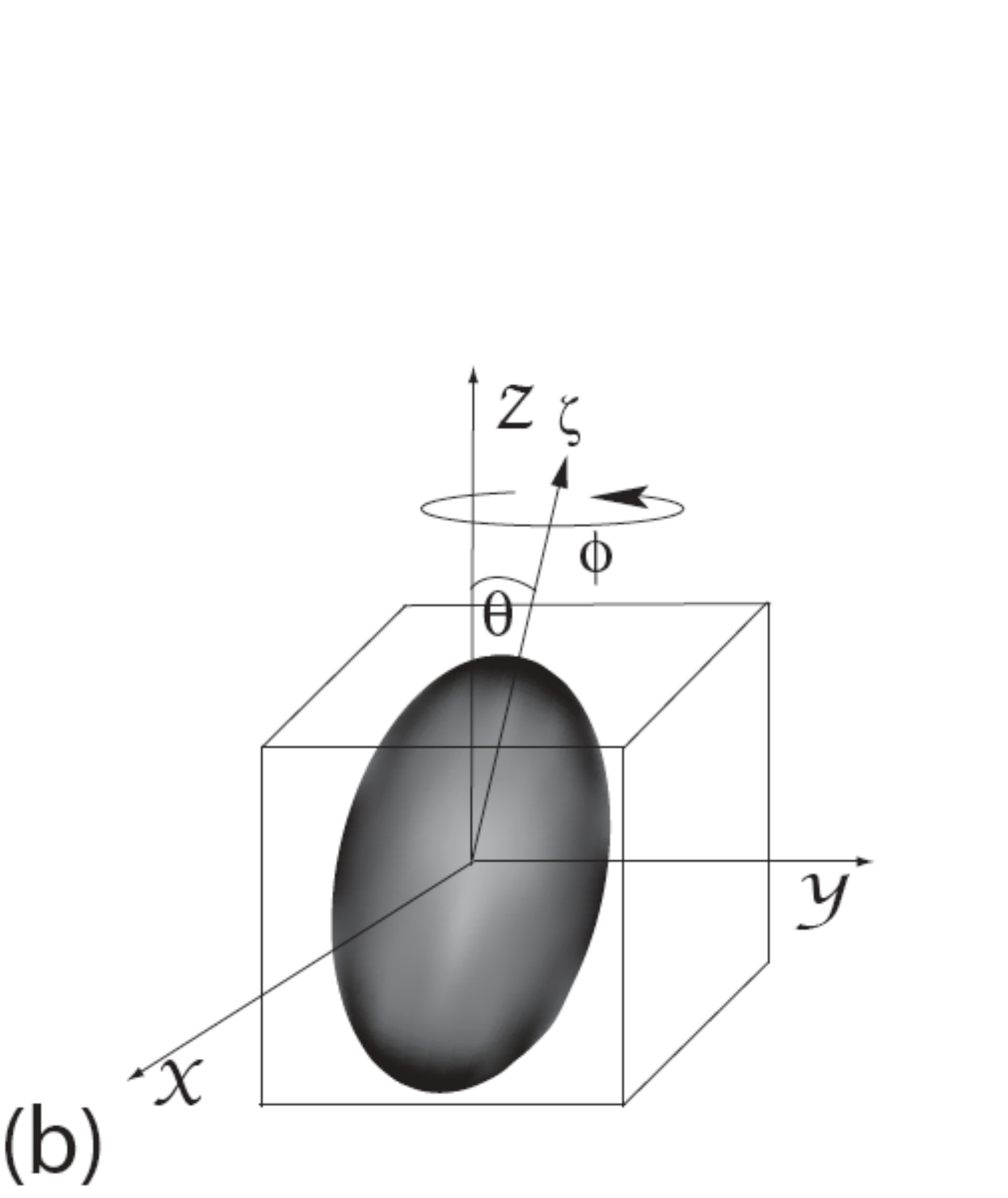}
    \end{tabular}
  \end{center}
 \caption{In a) an atom is in its ground state in a cell while a photon is incoming with circular polarization and momentum parallel to the axis of the cell. In b) the photon has been absorbed transferring its angular momentum to the atom, which precesses around the symmetry axis. A photon with momentum orthogonal to the cell axis cannot be absorbed, because the atom cannot absorb its angular momentum since it cannot rotate around any axis orthogonal to the cell axis.
  }
     \label{f2}
\end{figure}

In the first paper on crystals~[\onlinecite{Hata}]  it was proposed to search for scissors modes with a  resonance fluorescence experiment.  In such an experiment the dichroism   can help to eliminate the background. 

Then it was considered the case of ions characterized by Spin-Orbit Locking, namely ions in which the spin-orbit force is so strong to lock the total spin to the density profile in such a way that  they rotate together. For such a case a different experiment was proposed,  inspired by the experiments on Bose-Einstein condensates~[\onlinecite{Mara}].  In the latter ones  one induces oscillations of the condensate giving a sudden twist to the magnetic trap. In the case of ions with spin-orbit locking one induces oscillations of the ions in the cells  by an impulsive change of an external magnetic field and observes the photons emitted when the ion goes to the ground state.
 
 The estimate of the  energy of scissors modes  is affected by a large uncertainty, ranging from 10  to a few tens of eV. This makes the experiments difficult, because one have a spectrometer  covering a  large x-ray energy range with a resolving power  up to  $10^7$. Such a high value is due to the long lifetime of the scissors mode
  and could explain why, as far as we know, no hint of scissors modes shoved so far in the experimental measurements of spectra in crystals.

 We therefore propose  to excite scissors modes by an intense source of x-rays, and instead of detecting the photons emitted in the decay, to measure  the variation of the magnetic field associated with such  excitation. To this purpose we can   use a Superconducting Quantum Interference Device (SQUID), that  is suitable to measure extremely small magnetic fields.

The restoring force acting on the ions in our model is mainly due to the interaction of their electric quadrupole moment with the electric quadrupole moments of the ligands. Scissors modes will then be realized easier in compounds in which there are highly deformed  ions, and natural candidates are compounds containing  Rare Earths (RE). Moreover the $f$-electrons of the Rare Earths are compact and for this reason  they can be reasonably assimilated to a rigid rotor.

We  will consider  crystals and   gels in which Rare Earths are dispersed. The second case is very interesting, because the RE ions are well separated, so as to make  completely justified  our basic approximation that they  are non interacting with one another. It will turn out, unfortunately, that our type of experiment with gels looks  below the  limit of present  possibilities.

We  end this Introduction by noticing the many reasons of interest of Scissors Modes. On general grounds for every system they should affect the dispersive effects in the channels with their quantum numbers, which are $J^{\pi} = 1^+$. 
For given systems they provide specific pieces of information. In nuclear physics they are related to the superfluidity of deformed nuclei, in Bose-Einstein condensates provide a signature of superfluidity, in metal clusters they are predicted to be responsible for paramagnetism, in nanoparticles they give information about the moment of inertia of the macrospin. The knowledge of their properties should  be of some consequence also for crystals,  because  the magnetic anisotropy of these systems is at the origin of many interesting technological applications including magnetic storage devices and sensors, spin-torque nano-oscillators for high-speed spintronics and spin-optics~[\onlinecite{Smen}]. In particular one could establish or rule out the existence of Spin-Orbit-Locking in RE atoms.

We will start by describing, in Section II,  the ORM  that we will use to describe the dynamics of a deformed atom in a crystal cell or in a gel. Next we will discuss in Section III the estimate of the parameters entering our model. Finally  in Section IV we will evaluate the magnetic field generated by photon absorption to be measured by a SQUID and will present our conclusions in Section V.

\section{The One-Rotor-Model}

In the Single-Ion Model of magnetic anisotropy each rare earth ion is assumed to be independent from the others. In the study of the magnetic properties only the motion of the $4f$ electron system as a whole is considered, disregarding its excitations. In other words this system is treated as a rigid rotor with spin.  We introduce the principal frame of inertia of the ion with axes $\xi,\eta, \zeta$ and 
study its dynamics with respect to a frame of reference fixed with the cell with $x, y, z$-axes parallel to the cell axes. 
 We restrict ourselves to a rotor with axial symmetry and assume the symmetry axis parallel to the $\zeta$-axis. In quantum mechanics rotations around the symmetry axis are forbidden, and therefore the wave functions must satisfy the constraint
\be
 {\hat {\boldsymbol {\zeta}}}\,\cdot\,{\bf L} \, \psi = \,0 
\ee
where ${\bf L}$ is the angular momentum the ion. Its components in the cell frame become then those
of a point particle. 
 Assuming the symmetry axis of the ion along the $\zeta$-axis the kinetic energy operator  is
\begin{equation}
T=  \frac{\hbar^2}{2{ \mathcal I}} \left( - \frac{\partial^2}{\partial \theta^2} - \cot \theta  \frac{\partial}{\partial \theta}+
\frac{1}{\sin^2\theta} L_z^2 \right) +V    \label{Hrotor}
\end{equation}
where $\theta$ is the angle between the $z$,$\zeta$-axes,  $ L_z = - i  \hbar  \,{\partial \over \partial \phi}$ is the $z$-component of the orbital angular momentum, and ${\mathcal I}$ the moment of inertia   with respect to the $\xi$- and $\eta$-axes. For the potential we assume 
\be
 V = \frac{1}{2}C \sin^2 \theta 
\ee
where  $C$ is a restoring force constant.   

The scalar product for such a system remains defined by the measure
of integration over the Euler angles, which for functions independent
of $\gamma$ is 
\begin{equation}
  \<\,\psi_i\,|\,\psi_f\,\> = \int_{0}^{2\pi}d\varphi\int_{0}^{\pi}d\theta
  \sin\theta\,\psi_i^{*}(\theta,\varphi)\,\psi_f(\theta,\varphi)= \delta_{if}
\end{equation}
The exact eigenfunctions and eigenvalues are known [\onlinecite{Abra}],
but in the cases of physical interest the quadratic approximation is sufficient
\begin{eqnarray}
  T &=& \,\frac{\hbar^2}{2{\mathcal I}}\,\biggl(\,- \frac{\partial^2}{\partial\,\theta^2}
  - \frac{1}{\theta}\,\frac{\partial}{\partial\,\theta}
  + \frac{1}{\theta^2}\,L_{z}^2 \,\biggr), \quad 0\le \theta \le \pi \nonumber \\
  V &=& 
  \left\{ 
    \begin{array}{rl}
      &1/2\,C\,\theta^2 \hspace{12mm},\quad\,0\,\le\,\theta\,\le\,\pi/2,\\
      &1/2\,C\,(\,\pi - \theta\,)^2\,\,\,,\quad\,\pi/2\,<\,\theta\,\le\,\pi.
    \end{array}
  \right.
  \label{eqn:kin}
\end{eqnarray}
We recognize that this Hamiltonian, in each of the $\theta$-regions above,
is exactly that of a two-dimensional harmonic oscillator,
provided we identify $\theta$ with the polar radius. As shown by the
following estimates, the fall off of the wave function is so fast
that we can extend without any appreciable error the integral over $\theta$
up to infinity.

In the quadratic approximation, in the region $0\,\le\,\theta\,\le\,\pi/2$
the eigenfunctions are
\begin{eqnarray}
  \psi_{n,l_z}(\theta,\varphi) &=& \frac{1}{\sqrt{2\pi}}\,\,
  {e^{il_z\varphi}}\chi_{ n, |l_z| }(\theta)
\end{eqnarray}
where
\begin{eqnarray}
  \chi_{ n, |l_z| }(\theta) = \nu_{ n, |l_z| }\biggl(\frac{\theta}{\theta_{0}}\biggr)^{|l_z|}
  {\rm exp}\biggl(-\frac{\theta^2}{2\theta_{0}^2}\biggr)L^{(|l_z|)}_{n}
  \biggl(\frac{\theta^2}{\theta_{0}^2}\biggr).
\end{eqnarray}

$L^{(|l_z|)}_{n}$ are Laguerre polynomials  and

\begin{eqnarray}
  \nu_{ n, |l_z| }=\frac{1}{\theta_0} \sqrt{\frac{2 \, n!}{(\,n+|l_z|)!}},
 \, \,\,\,\theta_0^2=\frac{\hbar}{\sqrt{{\mathcal I}\,C}}.
\end{eqnarray}
The eigenvalues are
\begin{equation}
  E_{n,l_z} = \hbar\omega\,(\,2n + |l_z| +1\,)\,,\hspace{5mm}(\,n = 0,\,1,\,2\,...)
\end{equation}
where 
\begin{eqnarray}
  \omega = \sqrt{\frac{C}{{\mathcal I}}}.
\end{eqnarray}
The first excited states have $n=0$, $l_z=\pm 1$, and as we will see, they are the only states
strongly coupled to the ground state by electromagnetic radiation.
As shown in  Fig.2, they describe
the precession of the atom at an angle $\overline{\theta} \sim \theta_0$
around the axis of the cell. 

If the ion is spinless its  wave function is invariant under the inversion of the orientation of its symmetry axis, that is not observable. A discussion of the consequences of this symmetry can be found in [\onlinecite{Palu}], but they will be ignored here because  most RE ions do have a nonvanishing spin. 

The dynamics is determined by the magnitude of the zero point fluctuations $\theta_0$. When $\theta_0 \rightarrow 0 $  {\it the axis of the rotor can be assumed to lie along the $z$-axis, the direction of easy magnetization, and its zero-point fluctuations  can be ignored}. For $\theta_0 \sim 1$ the zero-point fluctuations cannot be neglected, but the rotor is still polarized within an angle of order $\theta_0$.  {\it For  $\theta_0 >> 1$   there is no polarization at all}. 

We quote the values of $\theta_0$ in some systems:   in the atomic nuclei~[\onlinecite{LoIu}] of  the rare earths, $\theta_0^2\approx 10^{-2}$;
in the crystal  LaMnO$_3$, $0.1< \theta_0^2 < 0.5$, depending on how the moment of inertia and restoring force constant are evaluated. Note that in the mentioned compound the deformed ion is Mn.

\section{Estimates of the parameters}

For numerical evaluations we need estimates of the moment of inertia ${\mathcal I}$ and of the restoring force constant  $C$ of the RE ions.  

 For the moment of inertia   we  should assume, consistently with our One-Rotor Model, the expression appropriate to a rigid body
\be    
    {\mathcal I}= \frac{2}{5}m_e Z\<r_{Z}^2\> \label{moment}
 \ee
where  $m_e$ is the electron mass, $\<r_{Z}^2\> $ the mean square radius of the electrons that contribute to the moment of inertia and $Z$ their number. We must then establish which are the electrons that  effectively take part in the rotation. Indeed one might include or exclude in the evaluation of ${\mathcal I}$ those electrons whose distribution has a spherical shape~[\onlinecite{Hata2}]. For a RE this means to include or exclude all the electrons of the inner shells and the $4f$-electrons that have a spherical density.  In the presence of Locking, however, the $4f$-electrons that  have a spherical distribution have a non-vanishing spin and therefore rotate with the electrons which determine the charge deformation, so that the ambiguity is restricted to the inner shells. If we assume that  only the $4f$-electrons participate in the rotation, we can set  $\<r_Z^2\> = \<r_{4f}^2\> \approx a_0^2$ 
 ($a_0 \approx 0.5 \, $\AA)  and $Z = Z_{4f} \approx 10$ if we consider, for instance, Dy, Ho and Er, getting
  \be
 \frac{\hbar^2}{ {\mathcal I}} \approx 8 \, eV.  \label{valuemoment}
  \ee  
  The main contribution to the restoring force comes from the interaction between the electric quadrupole moment of the RE ion and the electric quadrupole moment of the ligands. So the  expression of $C$ depends on the specific structure. One example of its calculation can be found in (\onlinecite{Hata}), but here we do not want to restrict ourselves to particular  cases. Therefore  we will avoid the evaluation of $C$ and parametrize the energy of the scissors modes according to
\be
 E_S= \hbar {\sqrt \frac{C}{{\mathcal I}}} =  \frac{\hbar^2}{{\mathcal I}}\frac{1}{\theta_0^2} > 8 \, \theta_0^{-2} eV\,. \label{energy}
\ee
We will assume that $\theta_0^2$ must be smaller than 1 if the RE ion must be polarized, and in specific cases we indeed found $\theta_0^2 \approx 0.5$.

It is also interesting to have an estimate of the resolution power  for our experiment. The  expression of the width of the scissors mode was determined in Refs. [\onlinecite{Hata1,Hata2}]
\be
\Gamma \approx  \frac{3}{4} \alpha \left( \frac{\hbar^2} {{\mathcal I}}\right)^3 \frac{1}{(m_e c^2)^2}\, 
\theta_0^{-8} > 1.6 \cdot 10^{-10} eV\,.
\ee
Notice the strong dependence on $\theta_0$.
With our parametrization of the energy we get
\be
\frac{E_S}{\Gamma}= \frac{4}{3 \alpha} \left( \frac{m_ec^2 {\mathcal I}}{\hbar^2} \right)^2 \theta_0^6 = 6 \cdot 10^8 \theta_0^6< 7 \cdot 10^7\,.\label{power}
\ee
We must notice that the above expression is not entirely consistent, because while the expression of the scissors mode energy has a general validity, the width was evaluated for a specific structure of the a crystal cell.
Nevertheless we think it should be sufficient to give an order of magnitude of the resolving power.

\section{Estimate of the magnetic field generated by photon absorption}

We  assume that all the crystal cells have their  easy axis parallel  to the z-axis of the laboratory frame. This situation is realized in single crystals and can be easily achieved by applying a convenient magnetic field. Otherwise an average over the cell axes should be performed as in [\onlinecite{Hata4}]. We also assume that the photons are circularly polarized (left or right is not relevant). The change of the magnetic moment of the target is given by the probability to excite a scissors mode times the scissors mode magnetic moment minus the ground state magnetic moment. It can be seen that the spin contribution is $\theta_0^4$ times smaller than the orbital one and therefore it will be neglected. The orbital magnetic moment in the ground state vanishes and then
 \be
\bigtriangleup \mu_z = w \, t_{flux} \, N_a <\psi_{0l_z}| \frac{1}{\hbar} L_z| \psi_{0l_z} >  \, \mu_B
\ee
where  $t_{flux}$ is the duration of the photon flux, $N_a$  the effective number of atoms in the sample and $\mu_B$ the Bohr magneton and
 $w$ is the photon absorption probability per unit time per unit frequency, that is given by
 \be
 w= 4 \pi^2 \alpha N(\omega) 
 \frac{\hbar^2 c}{m_e^2 \omega} |(M_{\epsilon}(\omega))_{fi}|^2\,.
 \ee
 In the above equation $m_e$ is the electron mass, $\alpha$ the fine structure constant, $c$ the velocity of light, $N(\omega)$ the number of photons of energy $\omega$ per unit volume and $M(\omega)$ the magnetic dipole moment matrix element
\be
\left(M_{\epsilon}\right)_{fi}= - \frac{\omega}{2 c}<\psi_f | \frac{1}{\hbar} L_{\epsilon}|\psi_i>
\ee
where
\be
L_{\epsilon} = - \frac{1}{{\sqrt 2}} \, e^{\pm i \phi}\left( \frac{\partial}{\partial \theta} + i \epsilon  \cot \theta  \frac{\partial}{\partial \phi}\right)
\ee
and $ \epsilon= \pm1 $ is the photon polarization.

Using the expression of the wave functions, for the quantum numbers $f=(0,m), i=(00)$ 
\be
\left(M_{\epsilon}\right)_{0l_z,00} = - \frac{i  \omega}{2 {\sqrt 2} \, c \,  \theta_0} \, \delta_{\epsilon, l_z}
\ee
\be
<\psi_{ol_z}|  \frac{1}{\hbar} L_z|\psi_{0l_z}> =l_z \,, \,\,\, l_z = \pm1
\ee
 so that
 \be
w= \frac{\pi^2}{2} \alpha \, \frac{1}{\theta_0^2}  \lambda_e^2  \frac{\omega}{\bigtriangleup \omega}\,  \frac{N_{phot} }{S_{source}}
\ee
 where  $\lambda_e$ is the Compton wave length of the electron, $\omega$ the photon frequency, $ \bigtriangleup \omega$ the frequency spread, $N_{phot} $ the number of incoming photons per unit time and $S_{source} $ the cross section of the photon beam. 

 The  effective number of atoms is given by
\be
N_a=\rho \, \lambda_{phot} \, S_{samp}
\ee
where $\rho$ is the number of atoms per unit volume, $S_{samp}$ the sample surface and $\lambda_{phot}$ the photon rms path. The above formula holds provided $\lambda_{phot}$ is smaller than the sample thickness.

We assume the bandwidth proportional to the photon frequency according to 
\be
 \bigtriangleup \, \omega = f \, \omega_{phot}= f \, \omega_{sciss}
\ee
so that finally 
\be
|\bigtriangleup \mu_z| = \frac{\pi^2}{2} \alpha \, \frac{1}{\theta_0^2} \frac{1}{f} \, N_{phot} \,t_{flux} \,  \frac{S_{samp}}{S_{source}}
 \lambda_e^2 \, \lambda_{phot} \, \rho \,  \mu_B\,.
\ee
A reasonable assumption is that
\be
S_{source} \approx S_{samp}\,, \,\,\,  f= 10^{-3}
\ee
so that
\be
|\bigtriangleup \mu_z| = \frac{\pi^2}{2} \alpha \, \frac{1}{\theta_0^2} \, N_{phot} \, t
\,  \lambda_e^2 \,  \lambda_{phot} \,\rho \, 10^{3}\, \mu_B\,.
\ee
The strength of the magnetic field along the z-axis generated by this variation of the magnetic moment  at a distance $r$ is
\be
B_z= \frac{\bigtriangleup \mu_z }{r^3}\mu_0
\ee
where $\mu_0$ is the vacuum magnetic permeability.

Next we assume
 \be
 N_{phot} \approx 10^{11} sec^{-1}
\ee
\be
 \theta_0^2 < 0.5.
 \ee
 To proceed further with numerical estimates we must distinguish the case of a dense RE compound sample as a crystal from the case of gels containing highly diluted RE atoms. 
 
 All the following numerical estimates will be done in S.I. units. Then 
  \be
   \mu_0= 1.25 \cdot 10^{-6} H/m\,, \,\,\, \mu_B= 9.3 \cdot 10^{-24} J/T
 \ee
 and
 \be
 \lambda_e=4 \cdot 10^{-13} m\,.
 \ee

\subsection{Dense matter}
 
We assume one RE atom per cell and a cell volume of $(3.5 A)^3= 4.3\cdot 10^{-29} \, m^3$, so that $\rho_{RE}= 2.3\cdot 10^{28 } \, m^{-3}$.
 
 We then have
 \be
 |\bigtriangleup \mu_z|= 7\cdot 10^{12} \, t_{flux} \, \lambda_e^2 \,  \lambda_{phot} \,\rho \, \mu_B
\ee 
and
\be
B= 7\cdot 10^{12} \, t_{flux} \, \lambda_e^2 \, \lambda_{phot} \,\rho \, r^{-3} \mu_0 \mu_B \,.
\ee
Setting
\be
\lambda_{phot}\sim 10^{-4} \, m\,, \,\,\,r \sim 10^{-2} \, m
\ee
we get
\be
B \sim 4 \cdot  \, 10^{-11} \, t_{flux}\, \mbox{T}\,, \,\, t_{flux} \mbox{ in seconds}\,.
\ee

Assuming an acquisition of one hour  we should measure a magnetic moment of the order of $10^{-7} \, \mbox{T}$, which is lower then the Earth's magnetic field on the surface whose magnitude ranges from 20 to 100  $\mu$T. However we can increase the signal  reducing the distance r of the detector and increasing the acquisition time and  the flux intensity. 

 Examples of suitable compounds for the proposed experiments are  R$_2$Fe$_{14}$B and  R$_2$Fe$_{17}$N$_3$ in which the RE ions  can be Dy, Ho and Er. We remind that the spin contribution to the variation of the magnetic moment is $\theta_0^4$ times smaller than the orbital contribution. Since all the quoted RE's  have  $Z_{4f} \approx 10$ and  $gJ \approx 10$, if $\theta_0$ is not sufficiently small the spin contribution cannot be neglected.
 
  Finally we remark  that the ions Gd$^{3+}$, La$^{3+}$, being spherical, do not have an orbital response. They can then be used to exclude, by comparison, spurious effects.

 \subsection{Soft matter}
 
 The difference with respect to a crystal is due to the difference of RE density and photon rms path
 \be
 B_{gel}= B_{crystal} \, \frac{(\rho_{RE})_{gel}}{ (\rho_{RE})_{crystal}}\,
 \frac{(\lambda_{phot})_{gel}}{ (\lambda_{phot})_{crystal}}
 \ee
 If $x$ is the percentage in weight of RE atoms in a gel made of $N_{gel}$ atoms of mass $m_{gel}$
 \be
 (\rho_{RE})_{gel} \, m_{RE}= x \, \rho_{gel} m_{gel}\,.
 \ee
Assuming $ m_{gel}/m_{RE} \sim 0.1$, $ \rho_{gel} \sim 0.5 \cdot 10^{21} \, m^{-3}$
\be
 (\rho_{RE})_{gel}  \approx x \cdot 0.5 \cdot 10^{-9} \left(\rho_{RE}\right)_{crystal}\,.
 \ee
With a  percentage in weight of x=0.05 we will have a magnetic moment 
$B \approx  10^{-21} \, t_{flux}$\, T\,. Considering that $10^{-15}$ \, T is the minimum value of the magnetic field actually detectable by existing SQUID devices we should imagine an acquisition of tens or even hundreds of hours or reduce significantly the distance and increase the flux intensity to be able to detect a signal from such a diluted RE system.

\section{Conclusion}

The  previous experiments designed to observe scissors modes in dense matter are affected by two orders of difficulty. The first one  is the detection of  photons in a wide energy spectrum, due to the uncertainty  in the theoretical value of the energy  of these excitations, that ranges between 10 and a few tens of eV. The second one is the need of a  resolving power of the order of $10^7$, much larger than that routinely used, that is of the order of $10^2 - 10^4$. These difficulties might also explain why, if scissors modes exist, they have not been noticed until now.

We propose here an experiment  in which, instead of detecting the decay photons,  one has to measure with a sensitive SQUID the magnetic field associated with  the excitation of scissors modes    induced  by photon absorption. For such an experiment we only need a sufficient flux of photons and a spectrum of the source sufficiently wide to cover the theoretical range of the scissors modes energy.

  There are some features of the present  experiment that should help to eliminate spurious signals

1) there should be no magnetic response with a source of linearly polarized photons

2) the intensity of the variation of the  magnetic field should decrease as the inverse of the cubic distance between the SQUID sensor and the sample

3) there should be no magnetic response from spherical RE ions.

4) because our calculations are performed assuming that the cells axes are aligned,  the induced variation of magnetic field should be proportional to the square of the cosine of the angle between the direction of the photons and the easy axes of the cells.  

Alignment  of the axes is realized in single crystals, and can also be achieved  by applying an external magnetic field. As far as RE dispersed in gels matrices are concerned, it appears possible to synthesize samples with a significant  alignment.

If the direction of the cells axes are disordered, we must perform an average as done in [\onlinecite{Hata4}] for nanoparticles. Obviously such an average does not change the order of magnitude of the results.

We must distinguish now between  dense compounds and light systems as RE ions dispersed in a gel matrix. 

In the first case  with the available photon fluxes in a synchrotron radiation facility in the ultraviolet  energy range,
we estimated  the  variation of the magnetic field  to be of the order of $10^{-7} \, \mbox{Tesla}$. This is to be compared with the typical precision of measurement of the terrestrial magnetic field that is of the order of one nanoTesla. 
 
In the second case, with  a small amount of RE ions dispersed in a gel matrix, our estimate yields  a magnetic field whose magnitude is well beyond the sensitivity of these superconducting devices. 
Our experiment in this case although unfeasible at present, is much more interesting from a theoretical point of view. Indeed for such dilute systems our basic assumption that the RE ions are independent from one another is completely justified and therefore the theoretical interpretation of the results is much cleaner.
Also  experiments on gels, however,  could become soon feasible improving the experimental layout e.g., with a smaller  device-sample distance, improving the SQUID  with a dedicated design, or in the next future using a free electron laser in order to have a photon flux in the ultraviolet energy  range  three to four orders of magnitude higher then that  of a storage ring.

We think that the successful detection of collective excitations modes such as scissors modes in RE compounds would trigger the search of a similar behavior in many other systems and, in particular, in actinides samples in which  paramagnetic and deformed atoms can be present in different concentrations.

\vskip2mm

\end{document}